\documentclass[12pt,a4paper]{article}
\usepackage[margin=1in]{geometry}
\usepackage{graphicx, setspace, amssymb, amsmath, amsthm, amstext, booktabs, float, multirow, subfigure, rotating, natbib, authblk, times}
 \pdfoutput=1 

 \newtheorem{thm}{Theorem}[section] \newtheorem{ass}{Assumption}[section] 
\newtheorem{lem}{Lemma}[section] \newtheorem*{prove}{Proof}
\begin{document}
\title{A Nonparametric Measure of Local Association for two-way Contingency Tables}
\author[1]{Francis K.C. Hui}
\author[1]{Gery Geenens\thanks{Corresponding author: ggeenens@unsw.edu.au; School of Mathematics and Statistics, The University of New South Wales, 2052, Sydney, Australia}}  
\affil[1]{School of Mathematics and Statistics, University of New South Wales, NSW 2052, Australia}
\date{}
\maketitle
\thispagestyle{empty} 

\begin{abstract}
In contingency table analysis, the odds ratio is a commonly applied measure used to summarize the degree of association between two categorical variables, say $R$ and $S$. Suppose now that for each individual in the table, a vector of continuous variables $X$ is also observed. It is then vital to analyze whether and how the degree of association varies with $X$. In this work, we extend the classical odds ratio to the conditional case, and develop nonparametric estimators of this ``pointwise odds ratio'' to summarize the strength of local association between $R$ and $S$ given $X$. To allow for maximum flexibility, we make this extension using kernel regression. We develop confidence intervals based on these nonparametric estimators. We demonstrate via simulation that our pointwise odds ratio estimators can outperform model-based counterparts from logistic regression and GAMs, without the need for a linearity or additivity assumption. Finally, we illustrate its application to a dataset of patients from an 
intensive care unit (ICU), offering a greater insight into how the association between survival of patients admitted for emergency versus elective reasons varies with the patients' ages. 
\end{abstract}

\section{Introduction} \label{sec:intro}
Consider a two-way contingency table with row and column variables $R$ and $S$, having levels $i = 1,\ldots,r$ and $j = 1,\ldots,s$ respectively. A commonly used measure to summarize the degree of association between $R$ and $S$ is the odds ratio. In the case of $r=s=2$, the odds ratio exhibits the simple form
\begin{equation} \label{eqn:globalOR}
OR = \frac{p_{11}p_{22}}{p_{12}p_{21}}
\end{equation}
where $p_{ij}=P(R=i,S=j)$. A sample estimate of $OR$ is obtained by replacing $p_{ij}$ with the observed sample proportions $\widehat{p}_{ij} = n_{ij}/n$. Due to its intuitive interpretation in terms of odds and conditional probabilities, $OR$ is often used in general $r \times s$ tables also, generating a set of odds ratios \citep[][Chapter 2]{agresti02}. \\
Suppose now that for each observation making up the table, a vector of continuous covariates $X$ is also observed. As a motivating example, we consider a dataset from \citet{hosmer00}, comprising 200 patients discharged from an adult intensive care unit (ICU). The data is cross-classified into survival status following hospital discharge (0 = Lived; 1 = Died) and type of admission into ICU (0 = Elective; 1 = Emergency), as shown in Table \ref{tab:application}. Along with these two variables, the age of each patient at the time of admission was also recorded. We are interested in seeing whether and how the association between survival and admission type varies according to age. More generally, we want to quantify the degree of local association between $R$ and $S$ conditional on $X=x$. \\
A traditional method for accomplishing this involves discretizing $X$ into several levels, and considering the odds ratio in each partial table \citep{ahrens06}. This technique however does not preserve the continuous nature of $X$ (age), resulting in a potential loss of information. A more commonly applied method is a model-based one, utilizing the odds ratio resulting from the logistic regression model below, 
\begin{equation}\label{eqn:logreg}
\log \left(\frac{p_i}{1-p_i}\right) = \beta_0 + \beta_1 x_i + \beta_2 r_i + \beta_3x_ir_i,
\end{equation}
where $p_i = P(S=1|R=r_i, X=x_i)$ is the conditional probability of `success' for the i$^{\text{th}}$ observation. The local odds ratio is then given by $OR(x) = e^{(\beta_2 + \beta_3x)}$. For a general $r \times s$ table, an extension can be made using polytomous response regression \citep[][Chapter 7]{agresti02}). \\
However, since these odds ratios are by-products of Generalized Linear Models \citep[GLMs,][]{mccullagh89}, they incur the problems associated with parametric regression. The logit linearity assumption means these measures lack flexibility and risk model mis-specification. For instance, it is clear from (\ref{eqn:logreg}) that there is an overly strict demand for the odds ratio to be increasing or decreasing in an exponential manner over $X$. \\
To introduce greater flexibility, a commonplace alternative is to utilize a Generalized Additive Models \citep[GAMs,][]{hastie90} instead:
\begin{equation} \label{eqn:gammodel}
\log \left(\frac{p_i}{1-p_i}\right) = \beta_0 + f_1(x_i) + \beta_1 r_i + r_i f_2(x_i),
\end{equation}
where $f_1(\cdot)$ and $f_2(\cdot)$ are two separate smoothers of $x$. Equivalently, (\ref{eqn:gammodel}) can also be regarded as a varying coefficient model \citep{hastie93}. In fact, this model nonparametrically fits two separate curves, one for each level of $r$, and the log odds ratio estimate is obtained from the difference of these two curves \citep{hastie90}. Using GAMs to estimate local odds ratios has been considered before by \citet{zhao96, figuerias01} amongst others, although their motivation stemmed from a regression context and thus considered $R$ as continuous also. \citet{cadarso05} proposed estimation of odds ratios using GAMs with unknown link functions, but their developments were again limited to $R$ continuous. Additionally, their simulations only considered datasets of size $n=1000$, meaning performance is not assessed for low to moderate sample sizes. \\
In contrast, as reflected in the ICU dataset example, our motivation arises from analyses of contingency tables. We seek a flexible measure of local association that is \emph{not} model-based in any sense. \\
In this paper, we propose a fully nonparametric measure of conditional association, formed by extending the global $OR$ to the local case. By exploiting the flexibility of kernel regression, our ``pointwise odds ratio'' permits a continuous $X$, while avoiding the hazards of model mis-specification. Using kernel regression to estimate the pointwise odds ratio was first suggested by \citet{geenens10}, although it was not explored in any depth there. This idea was also independently proposed by \citet{chen11}, although our work explores the problem much more thoroughly. Specifically, we propose adjusted estimators of the pointwise log odds ratio which have better statistical properties compared to a basic plug-in approach. We also develop confidence intervals for these new estimators. Applying these methods to the ICU dataset, we are able to gain a more nuanced view of the underlying relationships between age, type of admission, and survival status. 

\section{The pointwise odds ratio} \label{sec:pointwiseOR} 
The pointwise odds ratio is an intuitive extension of the global odds ratio defined in (\ref{eqn:globalOR}), formed using the conditional probabilities $p_{ij}(x)=P(R=i,S=j|X=x)$,
\begin{equation} \label{eqn:ORx}
OR(x) = \frac{p_{11}(x)p_{22}(x)}{p_{12}(x)p_{21}(x)} \quad \forall x \in S_X,
\end{equation}
for $r=s=2$. Equation (\ref{eqn:ORx}) can be broadened to produce a set of pointwise odds ratios for a general $r \times s$ table, but we restrict developments here to the simplest case. Also for simplicity here, we restrict attention to univariate $X$, with the developments in this work generalizable to the multivariate case. Evidently, $OR(x) \ge 0$, with $OR(x) = 1$ implying conditional independence of $R$ and $S$ at $X=x$. \\
For the developments in this paper, the following distributional assumption is made:
\begin{ass} \label{ass:multinom} The sample of observations can be described by $\{(X_k, Z_k)\}_{k=1}^{n}$, which form a sequence of i.i.d. replications of $(X,Z) \in S_X \times \{z \in \{0,1\}^{4}: \sum\limits_{ij=11}^{22} z^{ij} = 1\}$, a random vector such that $Z|X \sim \text{Multinomial}(1, p(X))$, where $p(x) = (p_{11}(x), p_{12}(x), p_{21}(x), p_{22}(x))^t$. \end{ass}
We use the shorthand $\sum\limits_{ij=11}^{22}$ to denote $\sum\limits_{i=1}^2\sum\limits_{j=1}^2$. Assumption \ref{ass:multinom} underlies most cross-sectional studies and surveys, as well as epidemiological studies consisting of a single cohort at baseline (see the ICU example). Qualitatively, Assumption \ref{ass:multinom} states that for each cell $(i,j)$, we observe a binary response vector coming from component $(ij)$ of each $Z_k$. Along with $X_k$, Nadaraya-Watson regression \citep[NW,][]{wand95} can be used to estimate $p_{ij}(x)$ for all four cells. This estimator is a sensible one to choose since, being a locally weighted average, it automatically guarantees estimated probabilities between 0 and 1, unlike local linear or P-Spline estimators for instance. It also ensures maximum flexibility in the estimation of $OR(x)$. For $ij = 11, \ldots, 22$, we have
\begin{equation} \label{eqn:phatij} 
\widehat{p}^h_{ij}(x) = \sum\limits_{k=1}^nW_h(x,X_k)Z^{ij}_k \quad \text{where} \quad W_h(x,X_k) = K \left(\dfrac{x-X_k}{h}\right)\Big / \sum\limits_{k=1}^nK \left(\dfrac{x- X_k}{h}\right)
\end{equation}
where $K(\cdot)$ and $h$ denote the kernel function and bandwidth respectively. For the latter, an optimal $h$ is obtained by minimizing the asymptotic mean integrated square error (AMISE) of $\widehat{p}^h_{ij}(x)$. Defining $\nu_0 = \int K^2(x) dx$ and $\kappa _2 = \int x^2 K(x)dx$, then from standard kernel regression theory \citep{wand95} we have
\begin{equation} \label{eqn:hopt}
h^{opt}_{ij} = \left(\frac{\nu_0\int_{S_X}\sigma^2_{ij}(x)dx}{\kappa_2^2\int_{S_X}b^2_{ij}(x)f(x)dx}\right)^{1/5}n^{-1/5}, 
\end{equation}
where $\sigma^2_{ij}(x) = p_{ij}(x)(1-p_{ij}(x))$, $b_{ij}(x) = p ^{''}_{ij}(x)/2 + p ^{'}_{ij}(x)f^{'}(x)/f(x)$, and $f(x)$ is the marginal density of $X$. Although the theory suggests that we should use four different bandwidths, one for each cell, it is argued in \citet[Section 2.3]{geenens10} that it is more appealing instead to use a single, common $h$ for all cells, and this is what we will do here. Then, it was showed in the same paper that, if $\lim_{n\rightarrow\infty}\sqrt{nh^5} = \lambda \; \text{with} \; 0 \le \lambda < \infty$ then
\begin{equation*} 
\sqrt{nh}(\widehat{p}^h_{ij}(x) - p_{ij}(x)) \xrightarrow{d} N\left(\kappa _2\lambda b_{ij}(x), \frac{\nu_{0}}{f(x)}p_{ij}(x)(1-p_{ij}(x))\right).
\end{equation*} 
For $h \sim n^{-1/5}$, as suggested by (\ref{eqn:hopt}), we have $\lambda > 0$ implying the distribution of $\sqrt{nh}(\widehat{p}^h_{ij}(x) - p_{ij}(x))$ is not asymptotically centered at 0. To deal with this undesired feature, we choose a sub-optimal bandwidth $h = o(n^{-1/5})$ (``undersmoothing'', the bias is asymptotically negligible and the Mean Squared Error is dominated by the variance) as suggested among others by \citet{hall92}. 
% We thus replace Assumption \ref{ass:hbandwidth1} by
% \begin{ass} \label{ass:hbandwidth2} The common bandwidth $h = h_n$ satisfies $nh^5 \rightarrow 0$ and $nh \rightarrow \infty$ as $n \rightarrow \infty$. \end{ass}
A common choice is to take $h \sim n^{-1/4}$, and indeed in this article our developments will be exposed with this order of $h$ in mind. Hence, the bias in the normality statement asymptotically vanishes and one obtains
$$ \sqrt{nh}(\widehat{p}^h_{ij}(x) - p_{ij}(x)) \xrightarrow{d} N\left(0, \frac{\nu_{0}}{f(x)}p_{ij}(x)(1-p_{ij}(x))\right). $$ 
Utilizing the conditional multinomial Assumption \ref{ass:multinom} and the Cram\'er-Wold device, a vectorial version is finally obtained for $\hat{p}^h(x)= (\hat{p}^h_{11}(x), \hat{p}^h_{12}(x), \hat{p}^h_{21}(x), \hat{p}^h_{22}(x))^t$:
\begin{equation} \label{eqn:phatnormalitynobias}
\sqrt{nh}(\widehat{p}^h(x) - p(x)) \xrightarrow{d} N\left(0, \frac{\nu_{0}}{f(x)}(diag(p(x)) - p(x)p(x)^{t})\right)
\end{equation}
where $diag(p(x))$ denotes a $4 \times 4$ diagonal matrix with elements equal to the components of $p(x)$. \\
Now, a simple plug-in estimator of $OR(x)$ is given by simple substitution of the NW conditional probabilities,
\begin{equation*} 
\widehat{OR}^h(x) = \frac{\widehat{p}^h_{11}(x)\widehat{p}^h_{22}(x)}{\widehat{p}^h_{12}(x)\widehat{p}^h_{21}(x)}.
\end{equation*}
Furthermore, an asymptotic $(1-\alpha)\%$ confidence interval (CI) for $\log(\widehat{OR}^h(x))$ can be obtained via the delta method on (\ref{eqn:phatnormalitynobias}):
\begin{equation} \label{eqn:logorhatCI}
\left(\log(\widehat{OR}^h(x)) \pm z_{1-\alpha/2} \sqrt{\frac{\nu_0}{nh\widehat{f}^h(x)}\sum\limits_{ij=11}^{rs}\frac{1}{\widehat{p}^h_{ij}(x)}} \right) 
\end{equation}
where $z_{1-\alpha/2}$ is the $(1-\alpha/2)$ quantile of the standard normal distribution, and $f(x)$ is estimated by the standard kernel density estimator
\begin{equation*}
\widehat{f}^h(x) = \frac{1}{nh}\sum\limits_{k=1}^n K \left(\frac{X_k-x}{h}\right).
\end{equation*}
For simplicity, we use the same kernel $K(\cdot)$ and bandwidth $h$ as in $\widehat{p}^h_{ij}(x)$, although this does not need to be the case.  

\section{An amended estimator for $\log(OR(x))$} 
\subsection{Motivation}
Although the plug-in estimator and associated CI are easy to calculate, they suffer from two major drawbacks. First, like its classical unconditional counterpart, $\log(\widehat{OR}^h(x))$ may be severely biased in finite samples. This is confirmed in the simulations of Section \ref{subsec:simcompareest}. Second, if one or more of the $\widehat{p}^h_{ij}(x)$'s are close to 0, then $\widehat{OR}^h(x)$ will either be close to 0 also or highly inflated. Since (\ref{eqn:logorhatCI}) has asymptotic variance proportional to $\sum\limits_{ij = 11}^{22}1/p_{ij}(x)$, a small value for one of the $\widehat{p}^h_{ij}(x)$'s also significantly enlarges the (estimated) variance, making confidence intervals of little use. To remedy these two problems, we propose adding a small deterministic value $\varepsilon(x) > 0$ to each $\widehat{p}^h_{ij}(x)$. This leads to an amended estimator
\begin{equation} \label{eqn:amendedest} 
\log(\widetilde{OR}^h(x)) = \log\left(\frac{(\widehat{p}^h_{11}(x)+\varepsilon(x))(\widehat{p}^h_{22}(x)+\varepsilon(x))}{(\widehat{p}^h_{12}(x)+\varepsilon(x))(\widehat{p}^h_{21}(x)+\varepsilon(x))}\right).
\end{equation}
We seek a value of $\varepsilon(x)$ for which $\log(\widetilde{OR}^h(x))$ has asymptotically smaller bias compared to $\log(\widehat{OR}^h(x))$. Although other methods of bias correcting an odds ratio estimator are available \citep[see for instance,][who use bootstrapping]{wang97}, these techniques are likely to produce similar statistical improvements compared to simply adding a small $\varepsilon(x)$, at the cost of greater computational intensity. Also, it is important to recognize that such an approach (adding a small deterministic value to each probability) has been taken before for $OR$. Specifically, we have the adjusted measure proposed by \citet{haldane55} 
\begin{equation} \label{eqn:globalORadjest}
\widehat{OR}_{adj} = \frac{(\widehat{p}_{11}+\frac{1}{2n})(\widehat{p}_{22}+\frac{1}{2n})}{(\widehat{p}_{12}+\frac{1}{2n})(\widehat{p}_{21}+\frac{1}{2n})}
\end{equation}
as a reduced bias estimator of $OR$. Furthermore, \citet{walter91} compared several estimators of $OR$, and found $\log(\widehat{OR}_{adj})$ perform well with regards to bias and mean squared error. The form of (\ref{eqn:globalORadjest}) is insightful not only because it is analogous to $\widetilde{OR}^h(x)$, but it shows that the adjustment made was $O(n^{-1})$ i.e., the variance rate of the parametric estimators $\widehat{p}_{ij}$. This suggests it might be appropriate to select $\varepsilon(x) \sim (nh)^{-1}$ in our nonparametric setting i.e., the variance rate of the kernel based estimators.

\subsection{Choosing $\varepsilon(x)$} \label{subsec:amendment}
By applying a number of Taylor expansions and utilizing some standard kernel regression theory results on the moments of the NW estimator, we derived a general expression for the bias of the amended estimator (\ref{eqn:amendedest}), see Appendix \ref{app:proofs} for relevant assumptions and proof. It turns out that
\begin{align} 
Bias(\log(\widetilde{OR}^h(x))) &= h^2\kappa _2\sum\limits_{ij=11}^{22}(-1)^{i+j}\left(\frac{b_{ij}(x)}{p_{ij}(x)}\right) \nonumber \\
&\quad + \, \left(\varepsilon(x)-\frac{\nu_0}{2nhf(x)}\right)\sum\limits_{ij=11}^{22}(-1)^{i+j}\left(\frac{1}{p_{ij}(x)}\right) \label{eqn:biaslogortilde} \\
& \quad + O(\varepsilon^2(x)) + O(h^2\varepsilon(x))+ o((nh)^{-1}) + O(\varepsilon^3(x)) + O(h^2\varepsilon^2(x)), \nonumber \end{align}
where $b_{ij}(x)$ is given below (\ref{eqn:hopt}). From this, we propose two possible values of $\varepsilon(x)$ which, along with the plug-in estimator (estimator I, $\varepsilon(x) = 0$), are summarized in Table \ref{tab:threeest}. The first one is just $\varepsilon(x) = \nu_0/(2nhf(x))$, evidently canceling out the second term in (\ref{eqn:biaslogortilde}). The second one attempts to balance the first term also. Note that, with $h \sim n^{-1/4}$ as we suggested in Section \ref{sec:pointwiseOR}, the amendment $\varepsilon(x) = \nu_0/(2nhf(x))$ (estimator II in Table \ref{tab:threeest}) only simplifies but does not explicitly reduce the asymptotic bias, as the first term in $h^2$ asymptotically dominates the second one in (\ref{eqn:biaslogortilde}). In fact, for this to provide a definite asymptotic bias reduction, we would need $h^2 = o((nh)^{-1})$, which in turn requires $h = o(n^{-1/3})$. Demanding such a bandwidth leads to a substantial amount of undersmoothing, to the extent that variance dominates and overwhelms any bias reduction achieved in the first place. This is to be avoided, and hence we maintain a reasonable amount of undersmoothing, driven by $h \sim n^{-1/4}$. \\
For $\varepsilon(x)$ to explicitly reduce the asymptotic bias in this case, we need $\varepsilon(x) \sim h^2$. With $h^4 = o((nh)^{-1})$ (which is the case with $h=o(n^{-1/5})$), one can then rearrange (\ref{eqn:biaslogortilde}) to produce the second, more involved amendment, see estimator III in Table \ref{tab:threeest}. We call the amended estimator using this second value of $\varepsilon(x)$ $\log(\overline{OR}^h(x))$, to distinguish it from the previous one. \\
Despite estimator III being one which actually produces an asymptotic bias reduction, we instead advocate the simpler amendment $\varepsilon(x) = \nu_0/(2nhf(x))$, and thus $\log(\widetilde{OR}^h(x))$ as the preferred estimator of the pointwise log odds ratio. The reasons for this are four-fold: 1) the adjustment $\varepsilon(x) = \nu_0/(2nhf(x))$ has a simple form and interpretation. Intuitively, it is a straight nonparametric analog of the $1/(2n)$ adjustment in (\ref{eqn:globalORadjest}); 2) the amendment $\varepsilon(x) = \nu_0/(2nhf(x))$ is very simple to compute. In contrast, to calculate $\log(\overline{OR}^h(x))$, one needs to estimate the bias terms $b_{ij}(x) = p ^{''}_{ij}(x)/2 + p ^{'}_{ij}(x)f^{'}(x)/f(x)$. This could be done by plugging in kernel estimates of the derivatives \citep{rodriguez99}, using local cubic smoothing \citep{fan96}, or via bootstrapping \citep{rodriguez93}, although all of these methods are challenging to implement; 3) Unlike with $\varepsilon(x) = \nu_0/(2nhf(x))$, there is no guarantee of $\overline{OR}^h(x) > 0$, especially after substituting in the unknown quantities; 4) we demonstrate empirically in Section \ref{subsec:simcompareest} that, in finite samples, $\log(\overline{OR}^h(x))$ and $\log(\widetilde{OR}^h(x))$ are similar with regards to bias, but the latter always has lower mean squared error (MSE). \\
It is interesting to point out that our discussion of choosing $\varepsilon(x)$ somewhat mirrors discussions regarding the two mainstream methods for dealing with bias in nonparametric regression procedures: undersmoothing \citep{hall92} and explicit bias correction \citep{neumann95}. In estimator III, one would be making an explicit bias correction, whereas adopting $\varepsilon(x) = \nu_0/(2nhf(x))$ with $h = o(n^{-1/3})$ is analogous to the approach of undersmoothing. By choosing estimator II but keeping $h \sim n^{-1/4}$, we actually promote a hybrid approach which balances the two. \\
As a final note, with the general expression for the bias given by (\ref{eqn:biaslogortilde}) and the asymptotic variance used for constructing (\ref{eqn:logorhatCI}), we can derive an expression for the AMISE of the plug-in estimator $\log(\widehat{OR}^h(x))$. From there, it can be seen that, for the purpose of estimating the pointwise log odds ratio, the asymptotic optimal bandwidth should be $h \sim n^{-1/5}$, same as the order of the optimal bandwidth when estimating the functions $p_{ij}$ themselves. This offers theoretical justification for using a \emph{single} undersmoothed bandwidth $h \sim n^{-1/4}$ all over.

\subsection{Simulation study 1 - Bias and mean squared error} \label{subsec:simcompareest}
\subsubsection{Design}
\begin{figure}[H] 
\centering \caption{\footnotesize True pointwise log odds ratio $\log(OR(x))$ as a function of $x$ for the three simulation models.} \label{fig:truesimor} \medskip
\includegraphics[height = 0.3\textheight]{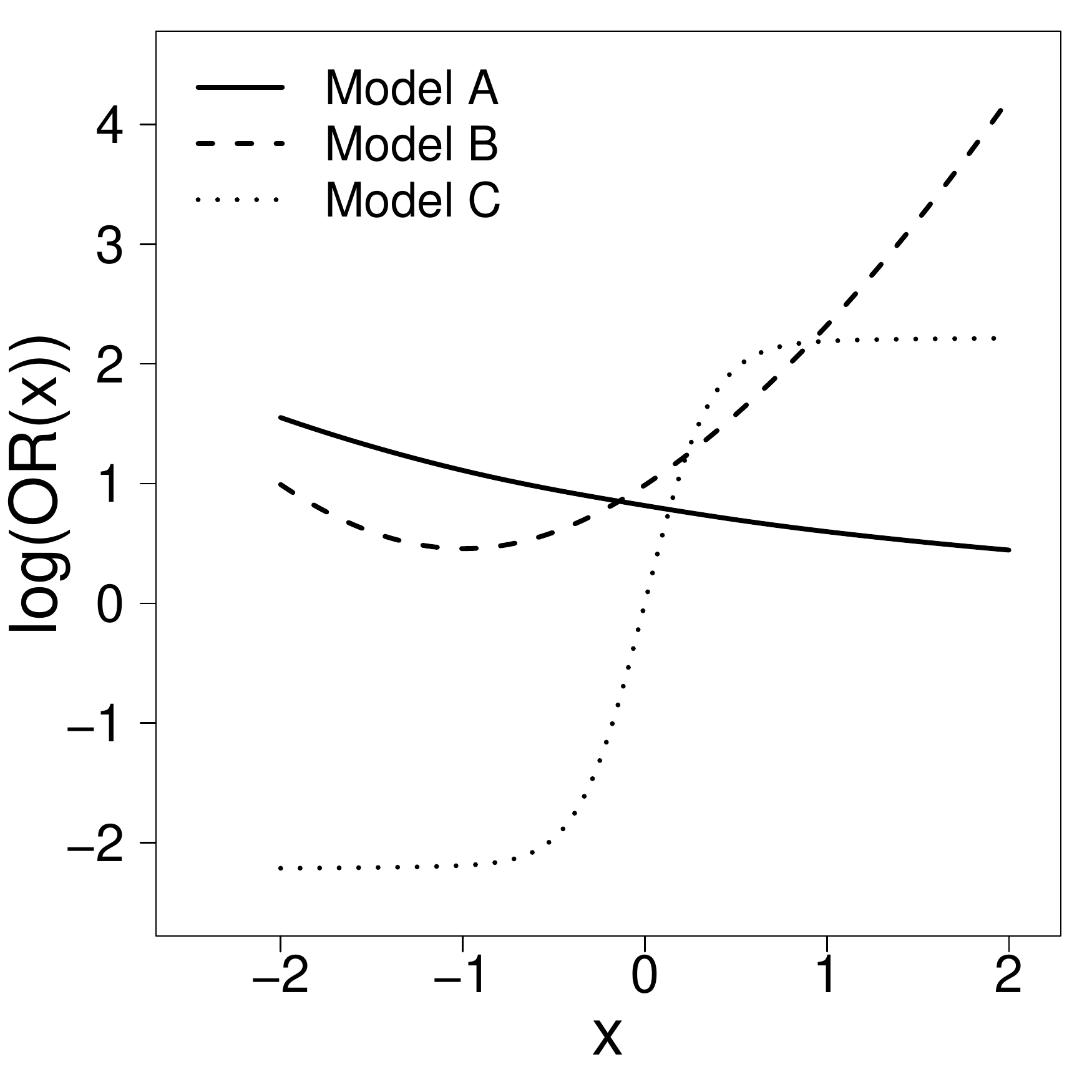} 
\end{figure}

We conduct a simulation study to compare the three estimators shown in Table \ref{tab:threeest} in terms of their bias and MSE. We also compare them to two model-based estimators: 1) an estimate of $\log(OR(x))$ based on the logistic regression of equation (\ref{eqn:logreg}), and given by $OR(x) = e^{(\beta_2 + \beta_3x)}$; 2) an estimate based on fitting the GAM model (\ref{eqn:gammodel}). The was done using the \texttt{mgcv} package in \texttt{R} \citep{wood06}.
%, using penalized regression splines with the penalty chosen via generalized cross validation (GCV). Although this type of smoother is somewhat different to kernel regression, it nevertheless serves the purpose of comparing statistical properties here. 
Three simulation models were designed:
\begin{align*}
X &\sim Unif[-2,2] \\
p_{1.}(x) = 0.07e^{-x^2} + 0.47  & \qquad  p_{.1}(x) = 0.1/(1+e^x) + 0.45 \\
p_{11}(x) = p_{1.}(x)p_{.1}(x) + \delta(x)  & \qquad  p_{12}(x) = p_{1.}(x)p_{.2}(x) - \delta(x)
 \end{align*}
with 
\begin{align*}
\text{Model A:} \; \delta(x) &= 0.05e^{-0.3x} \\ % mod2v2
\text{Model B:} \; \; \delta(x) &= 0.25 - \phi(x; -1, 1.8^2) \\ %mod2v3
\text{Model C:} \; \; \delta(x) &= 0.25\left((1+e^{-6x})^{-1} -0.5\right) %mod2
\end{align*}
where $\phi(x; \mu, \sigma^2)$ denotes the density of a normal distribution with mean $\mu$ and variance $\sigma^2$. The above design can be interpreted as follows: if $\delta(x) = 0$, then $p_{ij}(x) = p_{i.}(x)p_{.j}(x) \; \forall i,j = 1,2$ and thus $\log(OR(x)) = 0$. The delta function controls the degree of local association, as shown in Figure \ref{fig:truesimor}, which depicts the true log odds ratios curves for the three models. The design of our models, in particular our choices of $\delta(x)$, are such that the shapes of $\log(OR(x))$ are representative of commonly encountered non-linear relative risk functions in epidemiology \citep{zhao96}, whilst encompassing a realistic range of values. \\
We assessed performance using empirical integrated absolute bias and MSE, calculated by first working out the pointwise absolute bias and MSE in increments of 0.05 from $x = -1.75$ to $x = 1.75$, then averaging over all the increments. It is essential to take the pointwise absolute bias i.e., ignore the sign, so that when averaging to produce the integrated bias, these values do not cancel each other out due to symmetry. Also, even though the full support of $x$ is from -2 to 2, we limit ourselves to the interval (-1.75,1.75) to avoid boundary bias \citep{fan96}. Sample sizes $n = 50,100,250,1000$ were considered, with 4000 simulated datasets for each $n$. \\
For the nonparametric estimators I-III, a Gaussian kernel was used with bandwidth selected via direct plug-in \citep{ruppert95} plus ``manual'' undersmoothing (multiplying the optimal bandwidth by $n^{-1/20}$ so as to get a bandwidth proportional to $n^{-1/4}$, as it is commonly done). Strictly speaking, a Gaussian $K(\cdot)$ is not compactly supported on $[-1,1]$, although a slight technical argument can be included to make the results above hold for such choice \citep{collomb76}. For estimator III, the NW bias terms $b_{ij}(x)$ were estimated via the binary bootstrap \citep[][equation (6)]{rodriguez93}.

\subsubsection{Results}
In all models, estimator I performed poorly at the two smaller sample sizes (Tables \ref{tab:intbias}-\ref{tab:intmse}). By amending the estimated probabilities as in estimators II and III, the integrated bias was significantly reduced (Table \ref{tab:intbias}). Expectedly, estimator III produced the smallest integrated bias in most configurations, although estimator II also performed quite competitively. \\
A major problem suffered by estimator III was that sometimes the estimates of the odds ratio turned out to be negative. In Model B at $n = 50$, there were 1361 cases (out of $71 \times 4000 = 284,000$) where $\overline{OR}^h(x) < 0$. This occurrence of negative values was not resolved at larger sample sizes e.g., in Model C at $n = 1000$, there remained 105 cases of invalid estimates. In contrast, estimator II cannot suffer from this problem, obviously. \\
The shape of the true log odds ratio curves in Models B-C (see Figure \ref{fig:truesimor}) meant there was a clear mis-specification of mean structure in fitting (\ref{eqn:logreg}). Therefore, the GLM-based estimator suffered from inflated bias even at large $n$ (Table \ref{tab:intbias}). In contrast, the flexibility of kernel regression allowed estimators II and III to perform much better than its parametric counterpart. The performance of the GAM-based estimator was somewhere in between the GLM model and the kernel-based estimators II and III. This is expected, given the `hybrid' nature of the GAM-based estimator between the purely linear-logistic expression in (\ref{eqn:logreg}) and the entirely nonparametric kernel-based methods. \\
Although its bias was higher compared to estimators II and III, the GLM-based estimator performed best with regards to MSE in Model A. We found however that this was largely due to the inadequacy of using the direct plug-in method \citep{ruppert95} to select the bandwidth for NW regression. For relatively flat functions like Model A, direct plug-in often leads to significant undersmoothing \citep{signorini04}. To investigate this, we re-calculated nonparametric estimators I-III in Model A, using the same 4000 simulated datasets at each $n$, but this time estimating $h$ via cross-validation \citep{hardle85}. Results showed that for all three nonparametric estimators, there was a sizable decrease in integrated MSE (see Supplementary Material). Moreover, the decrease is such that estimator II actually had a lower integrated MSE than both the logistic regression and GAM estimators at all four sample sizes. Comparing cross-validation and direct plug-in, we found that the average $h$ based on the former was roughly five times larger than for the latter. \\
For Models B-C, estimator II had the lowest MSE for all sample sizes (Table \ref{tab:intmse}). Although estimator III marginally outperformed II with regards to integrated bias (Table \ref{tab:intbias}), the complexity and additional variability resulting from $\log(\overline{OR}^h(x))$ meant that it was the latter which had the lower MSE. \\
In conclusion, the simulation results presented here lead us to recommend using $\log(\widetilde{OR}^h(x))$ as a preferred estimator of the pointwise log odds ratio. Unless stated otherwise, future references to $\varepsilon(x)$ will admit the definition $\varepsilon(x) = \nu_0/(2nhf(x))$ only. 

\section{Confidence intervals} \label{sec:ci}
Given the strategy of adding that small value $\varepsilon(x)$ to the conditional probabilities, a first attempt at constructing confidence limits based on $\log(\widetilde{OR}^h(x))$ would be to adjust (\ref{eqn:logorhatCI}) in an analogous manner,
\begin{equation} \label{eqn:logortildeCI}
\left(\log(\widetilde{OR}^h(x)) \pm z_{1-\alpha/2} \sqrt{\frac{\nu_0}{nh\widehat{f}^h(x)}\sum\limits_{ij=11}^{rs}\frac{1}{\widehat{p}^h_{ij}(x) + \frac{\nu_0}{2nh\widehat{f}^h(x)}}}\right).
\end{equation}
The form above is simple to work with, and parallels the variance formula discussed in \citet[Section 3.1.1]{agresti02} for $\log(\widehat{OR}_{adj})$ in  (\ref{eqn:globalORadjest}). However, although we expect this to work better than (\ref{eqn:logorhatCI}), the use of resampling methods may offer even further improvements on this asymptotic result in regards to coverage probability and/or interval width \citep{horowitz01}. Therefore we explore this below. We also recognize that the delta method could have been applied directly to $\log(\widetilde{OR}^h(x))$, but we found this led to a very complex formula for the asymptotic variance, and so have avoided it here. \\
To obtain bootstrap based confidence intervals, we propose a new resampling procedure called the multinomial-1 bootstrap, inspired by some ideas in \citet{rodriguez93} and developed in \citet{hui12}. \\
Consider cell $(i,j)$ in our $2 \times 2$ table, for which we have a binary response $Z^{ij}_k$ and its corresponding covariate $X_k$, $k = 1,\ldots,n$. For resampling methods to work here, two requirements need to be satisfied: 1) the bootstrapped response variables $Z^{*ij}$ must be binary and satisfy $\sum\limits_{ij=11}^{22}Z^{*ij} = 1$; 2) we must capture the conditional nature of the probabilities ${p}_{ij}(x) = P(Z^{ij}|X=x)$. The multinomial-1 bootstrap therefore works by the following: first, estimate $p_{ij}(x)$ with (\ref{eqn:phatij}) using a pilot bandwidth $g$ (instead of $h$) to obtain the vector $ \widehat{p}^g(x) = (\widehat{p}^g_{11}(x), \widehat{p}^g_{12}(x), \widehat{p}^g_{21}(x), \widehat{p}^g_{22}(x))^t$. Then for $k = 1,\ldots,n$, we simulate a bootstrap response vector $Z_k^* = (Z_k^{*11}, Z_k^{*12}, Z_k^{*21}, Z_k^{*22})^t$ from
$$ Z^{*}_k \sim \text{Multinomial}\left(1,\widehat{p}^g(X_k)\right). $$
Having obtained the bootstrap sample $(X_k, Z^*_k)$, we re-perform kernel regression using the previous $h \sim n^{-1/4}$ to obtain $\widehat{p}^{\,*h}_{ij}(x)$ and hence the vector $\widehat{p}^{\,*h}(x)$. Use of an initial oversmoothed $g$ is typical when bootstrap is used in nonparametric regression \citep[see for instance,][]{hardle91}, and is necessary to properly account for the bias inherent in kernel regression. A pilot bandwidth $g \sim n^{-1/9}$ has been proved to be optimal in that purpose, and this is also what we will use in this work. By extending the theory of \citet{rodriguez93}, it may be shown that the multinomial-1 bootstrap produces a consistent estimator of $\log(\widetilde{OR}^h(x))$ (see Appendix \ref{app:proofs}). Percentile bootstrap confidence intervals based on $\log(OR(x))$ are thus obtained by generating a sufficiently large number of bootstrapped datasets, and calculating $(\alpha/2)$ and $(1-\alpha/2)$ quantiles of $\left(\log(\widetilde{OR}^{*h}(x)) - \log(\widetilde{OR}^g(x))\right)$. Denoting these quantiles by $l^*(x)$ and $u^*(x)$ respectively, a $100(1-\alpha/2)\%$ bootstrap confidence interval for $\log(OR(x))$ is given by
\begin{equation} \label{eqn:bootlogortildeCI}
\left(\log(\widetilde{OR}^{h}(x)) - u^*(x), \; \log(\widetilde{OR}^{h}(x)) - l^*(x)\right).
\end{equation}

\subsection{Simulation study 2 - coverage probabilities} \label{subsec:ecp}
We compare the three confidence intervals for $\log(OR(x))$, as represented by (\ref{eqn:logorhatCI}), (\ref{eqn:logortildeCI}) and (\ref{eqn:bootlogortildeCI}), in terms of their empirical coverage probability (ECP) and mean length (on a log scale). ECP is defined as the number of times the true pointwise log odds ratio lies within the generated CIs (nominated level 95\%), divided by the total number of replications. We used Models A-C established in Section \ref{subsec:simcompareest}, with $n = 50,100,250$ and 1000 simulated datasets for each $n$. CIs were calculated at values $x = -1,0,1.5$. For the bootstrap CIs, we used $B=500$ resamples. Initially we tested $B = 1000$, but found 500 replications produced similar intervals. The results are shown in Table \ref{tab:ecp}. \\ 
For all three models, the delta method procedure based on the plug-in estimator I (DM-I) lead to conservative CIs i.e., high ECP for $n=50$ and 100. DM-I also had the widest confidence intervals for all sample sizes. Such wide intervals (on a log-scale) will be of little use to the applied researcher when attempting to determine a realistic range of values for the true $OR(x)$. \\
Applying the delta method to $\log(\widetilde{OR}^h(x))$ (DM-II) lead to CIs with much smaller interval lengths, without any consistent decrease in ECP. The bootstrap percentile CIs (M1B-II) performed best, having almost always the smallest interval lengths with similar ECP. For locations where $\log(OR(x))$ was substantially different from 0 e.g., Model B $x = 1.5$ and Model C at $x = -1, 1.5$, bootstrap based intervals offered useful decreases in average CI length without being further away from the nominated 95\% coverage probability. Specifically, while the delta method intervals tended to have ECP $>$ 95\%, the bootstrap CIs often have coverage slightly below 95\%. This could be blamed in the name of conservatism, however, the \emph{absolute} deviations from the targeted level 95\% were very similar between the two methods. For $n = 250$, both DM-II and M1B-II performed equally well with regards to ECP and interval width.  

\section{A real-data application} \label{sec:applied}
\begin{figure}[H] 
\caption{\footnotesize Local log odds ratio of death for patients admitted to ICU for emergency reasons relative to those admitted for elective reasons, plotted against age. Plotted are the estimates from a logistic regression model fitted with an interaction effect (solid line), a GAM fit (dotted line), and pointwise log odds ratio based on estimator II $\log(\widetilde{OR}^h(x))$ (dashed line). A horizontal line at $\log(OR(x)) = 0$  marks local independence.} \label{fig:icu} \medskip \centering 
\includegraphics[height = 0.3\textheight]{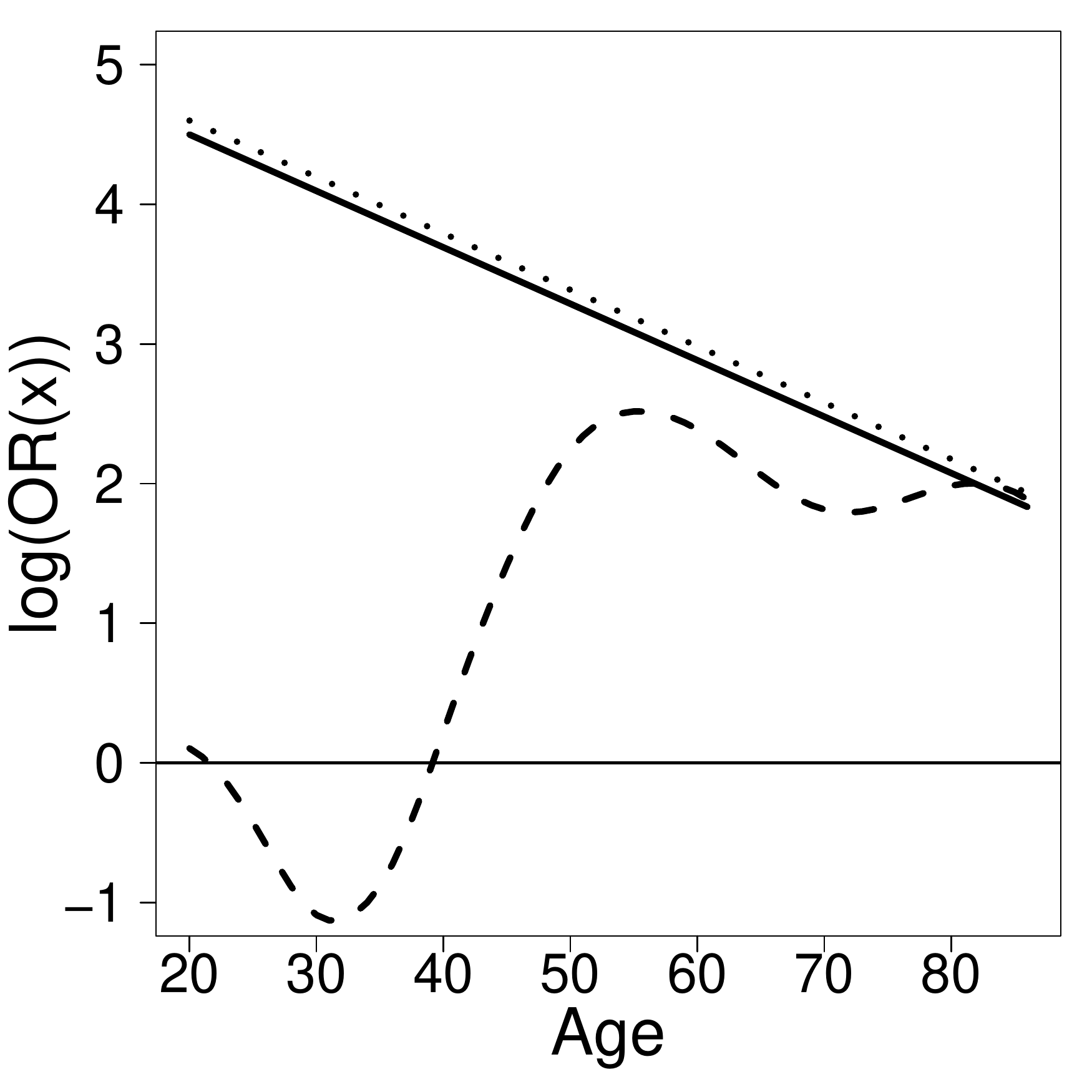}
\end{figure}

We illustrate the application of the methods developed to the ICU dataset discussed in Section \ref{sec:intro}. We are interested in exploring how the strength and  direction of the association between patient survival following hospital discharge and type of admission varies with the age of the patients. To begin, a Pearson $\chi^2$ test on Table \ref{tab:application} provided strong evidence against global independence ($p$-value = 0.001), and the global odds ratio estimate $\widehat{OR} = 8.89$ indicated that the odds of dying from an emergency admission was almost 9 times that for an elective admission, and could be as high as 38 times (95\% Wald CI: [2.064;38.290]). Although this conclusion is expected, it should be subject to further investigation, particularly in light of the hypothesis that the strength of this association may be weaker for young adults. \\
We first approached this investigation using logistic regression, with results indicating the main effect of admission type was significant given age ($p$-value $<$ 0.001). The interaction term between age and admission type however was not significant in this model ($p$-value = 0.622), meaning the odds ratio, despite being significantly greater than 1 ($e^{2.983} \approx 19.747$), did not appear to vary with age. Persisting with the interaction model, the log odds ratio estimate actually shows a decline with increasing age (Figure \ref{fig:icu} - solid line). We also fitted a GAM model, with penalized regression splines and penalty chosen via GCV, using the ``\texttt{by}'' argument available in the \texttt{mgcv} package \citep{wood06}. The resulting log odds ratio curve closely follows the fit from logistic regression (Figure \ref{fig:icu} - dotted line). \\
As an alternative to model-based approaches, we decided to use the pointwise log odds ratio estimated using $\log(\widetilde{OR}^h(x))$ (estimator II in Table \ref{tab:threeest}). The result plotted as the dashed curve in Figure \ref{fig:icu}. From ages 50 and 86, $\log(\widetilde{OR}^h(x))$ hovered around 2.5 which, in reasonable agreement with logistic regression, provided strong evidence for the odds of death for patients discharged from an emergency admission being significantly higher than those released from elective admission. However, for ages less than 50, $\log(\widetilde{OR}^h(x))$ drops to become non-significant. This is in contrast to both the logistic regression and GAM models which were not able to provide any notion of this dampening. \\
To further verify whether this decrease is substantiated, 95\% pointwise bootstrap confidence intervals ($B = 1000$) were calculated at ages 30, 50 and 70. At both ages 50 (CI: [1.607;4.157]) and 70 (CI: [0.260;3.076]) the limits were above 0, and confirmed that for older patients the odds of death was significantly higher for patients admitted for emergency reasons. However, for age 30 (CI: [-1.633;1.394]) the confidence interval contains $\log(OR(x)) = 0$, and indicated that for younger patients, there is no strong evidence to suggest type of admission into ICU affects the odds of survival. 

\section{Concluding remarks} \label{sec:conclude} 
In this paper, we developed a new measure of local association by extending the standard odds ratios using conditional probabilities, and estimating these probabilities nonparametrically using kernel regression to allow maximum flexibility. Three estimators of $\log(OR(x))$ were proposed, from which we recommend the amended estimator $\log(\widetilde{OR}^h(x))$, which is both simple to calculate and has good bias/MSE properties. We formulated confidence intervals based on $\log(\widetilde{OR}^h(x))$ using both asymptotic arguments and an innovative multinomial-1 bootstrap procedure. \\
One particular issue we did not explore is bandwidth selection for our estimators of $\log(OR(x))$. For kernel regression in general, there is no single best method of selecting the bandwidth. The direct plug-in method tends to perform well for estimating the functions $p_{ij}$ in practice in many cases \citep{signorini04, ruppert95}, which is why we chose it for this work. However, there is no real guarantee that it would perform as well for estimating our pointwise log odds ratio. Consequently, further studies need to be conducted evaluating various approaches of choosing $h$ in this very setting. Indeed, the results from the first simulation in Section \ref{subsec:simcompareest} provide clear evidence that a thorough comparison of the various methods in selecting $h$ is necessary. \\ 
In the future, we hope to develop model-free nonparametric association measures beyond the pointwise odds ratio e.g., pointwise relative risk, pointwise Kendall's tau and so on. How confidence intervals can be established for these quantities is also of interest. Finally, the use of kernel regression means that due to the curse of dimensionality, it is inefficient to produce a pointwise odds ratio which is `local' with respect to many covariates. Perhaps the use of semi-parametric methods e.g., single index models, to estimate the conditional probabilities instead can overcome this problem.

\section*{Acknowledgements}
FH was supported by a Faculty of Science Honours Scholarship at the University of New South Wales. GG was supported by a Faculty Research Grant from the Faculty of Science, University of New South Wales.

%-------------------------------------------------------------

%---------------------------------------------
\newpage 
\section*{Tables and Figures} 
\begin{table}[H] 
\caption{\footnotesize Dataset of 200 patients discharged from an adult ICU, classified according to survival status and type of admission.} \label{tab:application} 
\medskip \centering 
\begin{tabular}{lccc}
\toprule[1.5pt]
 & & \multicolumn{2}{c}{Status} \\
 & & \emph{Died} & \emph{Lived} \\
\cmidrule{3-4}
\multirow{2}{*}{Admission} & \emph{Emergency} & 38 & 109 \\
 & \emph{Elective} & 2 & 51 \\
\bottomrule[1.5pt] 
\end{tabular} \end{table}

\begin{table}[H] 
\caption{\footnotesize Summary of the three kernel based estimators for the pointwise log odds ratio proposed in this work.} \label{tab:threeest} 
\medskip \centering 
\begin{tabular}{cll} 
\toprule[1.5pt]
Estimator & Notation & Amendment \\
\midrule
I & $\log(\widehat{OR}^h(x)) \qquad$ & $\varepsilon(x) = 0 \quad \forall x \in S_X$ \\
II & $\log(\widetilde{OR}^h(x)) \qquad$ & $\varepsilon(x) = \nu_0/(2nhf(x))$ \\
III & $\log(\overline{OR}^h(x))\qquad $ & $\varepsilon(x) = \frac{\nu_0}{2nhf(x)} - h^2\kappa _2\left(\frac{\sum\limits_{ij=11}^{22}(-1)^{i+j}\left(\frac{b_{ij}(x)}{p_{ij}(x)}\right)}{\sum\limits_{ij=11}^{22}(-1)^{i+j}\left(\frac{1}{p_{ij}(x)}\right)}\right)$ \\
\bottomrule[1.5pt] 
\end{tabular} \end{table}

\begin{table}[H] 
\caption{\footnotesize Integrated absolute bias of the three nonparametric (I,II and III), the GLM, and the GAM estimators of $log(OR(x))$ for Models A-C at various samples sizes $n$. The best estimator is each configuration is highlighted in bold.} \medskip \centering \label{tab:intbias}
\begin{tabular}{lcccccc}
\toprule[1.5pt]
Model & $n$ & I & II & III & GLM & GAM \\ 
\cmidrule{3-7}
\multirow{4}{*}{A} & 50 & 0.465 & 0.036 & \textbf{0.016} & 0.102 & 0.365 \\
 & 100 & 0.132 & \textbf{0.014} & 0.027 & 0.065 & 0.096 \\
 & 250 & 0.036 & \textbf{0.001} & 0.012 & 0.044 & 0.030 \\
 & 1000 & 0.012 & \textbf{0.006} & 0.006 & 0.040 & 0.011 \\
\cmidrule{3-7}
\multirow{4}{*}{B} & 50 & 1.200 & 0.111 & \textbf{0.094} & 0.408 & 0.463 \\
 & 100 & 0.316 & 0.056 & \textbf{0.046} & 0.364 & 0.218 \\
 & 250 & 0.075 & \textbf{0.024} & 0.027 & 0.357 & 0.082 \\
 & 1000 & 0.034 & 0.020 & \textbf{0.020} & 0.354 & 0.053 \\
\cmidrule{3-7}
\multirow{4}{*}{C} & 50 & 0.879 & 0.212 & \textbf{0.194} & 0.637 & 0.541 \\
 & 100 & 0.291 & 0.113 & \textbf{0.100} & 0.608 & 0.415 \\
 & 250 & 0.124 & 0.093 & \textbf{0.088} & 0.607 & 0.264 \\
 & 1000 & 0.076 & 0.066 & \textbf{0.061} & 0.607 & 0.163 \\
\bottomrule[1.5pt]
\end{tabular} \end{table}

\begin{table}[H] 
\caption{\footnotesize Integrated MSE of the three nonparametric (I,II and III), the GLM, and the GAM estimators of $log(OR(x))$ for Models A-C at various samples sizes $n$. The best estimator is each configuration is highlighted in bold.} \medskip \centering \label{tab:intmse}
\begin{tabular}{lcccccc}
\toprule[1.5pt]
Model & $n$ & I & II & III & GLM & GAM \\ 
\cmidrule{3-7}
\multirow{4}{*}{A} & 50 & 6.152 & 1.265 & 1.386 & \textbf{0.893} & 1.076 \\
 & 100 & 1.106 & 0.730 & 0.792 & \textbf{0.344} & 0.646 \\
 & 250 & 0.323 & 0.293 & 0.309 & \textbf{0.127} & 0.197 \\
 & 1000 & 0.077 & 0.075 & 0.079 & \textbf{0.042} & 0.045 \\
\cmidrule{3-7}
\multirow{4}{*}{B} & 50 & 17.496 & \textbf{1.209} & 1.308 & 1.238 & 1.863 \\
 & 100 & 2.429 & \textbf{0.748} & 0.809 & 0.851 & 1.116 \\
 & 250 & 0.398 & \textbf{0.292} & 0.311 & 0.300 & 0.293 \\
 & 1000 & 0.096 & \textbf{0.053} & 0.057 & 0.192 & 0.058 \\
\cmidrule{3-7}
\multirow{4}{*}{C} & 50 & 11.516 & \textbf{1.311} & 1.467 & 2.117 & 2.041 \\
 & 100 & 2.062 & \textbf{0.828} & 0.953 & 0.963 & 1.263 \\
 & 250 & 0.459 & \textbf{0.350} & 0.428 & 0.649 & 0.428 \\
 & 1000 & 0.132 & \textbf{0.079} & 0.130 & 0.534 & 0.246 \\
\bottomrule[1.5pt]
\end{tabular} \end{table}

\begin{sidewaystable} \footnotesize
\caption{\footnotesize Empirical coverage probability and mean widths of $95\%$ confidence intervals for $\log(OR(x))$ at various combinations of $n$ and $x$, based on Models A-C. Three methods were compared: Delta method using plug-in estimator I $\log(\widehat{OR}^h(x))$ (DM-I), delta method using amended estimator II (DM-II), and the multinomial-1 bootstrap percentile CIs based on amended estimator II (M1B-II). Results are presented in the format: ECP (mean width). } \label{tab:ecp} \medskip \centering 
\begin{tabular}{lcccccccccccc} 
\toprule[1.5pt]
 & & \multicolumn{3}{c}{Model A} & & \multicolumn{3}{c}{Model B} & & \multicolumn{3}{c}{Model C}\\ 
$x$ & $n$ & DM-I & DM-II & M1B-II & & DM-I & DM-II & M1B-II & & DM-I & DM-II & M1B-II\\
\cmidrule{3-13}
\multirow{4}{*}{-1} & 50 & 0.978 (6.51) & 0.976 (4.65) & 0.944 (4.47) & & 0.978 (6.87) & 0.974 (4.76) & 0.924 (4.26) & & 0.980 (9.34) & 0.969 (5.20) & 0.924 (4.18) \\
 & 100 & 0.972 (3.73) & 0.980 (3.28) & 0.960 (3.29) & & 0.980 (3.85) & 0.972 (3.40) & 0.956 (3.35) & & 0.984 (5.11) & 0.960 (3.85) & 0.940 (3.49) \\
 & 250 & 0.952 (2.09) & 0.952 (2.04) & 0.952 (2.01) & & 0.952 (2.19) & 0.954 (2.14) & 0.946 (2.14) & & 0.948 (2.55) & 0.942 (2.44) & 0.944 (2.35) \\\\
\multirow{4}{*}{0} & 50 & 0.988 (6.80) & 0.984 (4.69) & 0.950 (4.45) & & 0.986 (6.32) & 0.978 (4.77) & 0.944 (4.26) & & 0.988 (6.03) & 0.988 (4.64) & 0.942 (4.23) \\
 & 100 & 0.968 (3.69) & 0.970 (3.38) & 0.950 (3.34) & & 0.972 (3.62) & 0.974 (3.33) & 0.956 (3.31) & & 0.952 (3.59) & 0.956 (3.31) & 0.954 (3.28) \\
 & 250 & 0.950 (2.16) & 0.952 (2.11) & 0.954 (2.10) & & 0.958 (2.15) & 0.964 (2.10) & 0.948 (2.09) & & 0.966 (2.15) & 0.972 (2.10) & 0.954 (2.09) \\\\
\multirow{4}{*}{1.5} & 50 & 0.968 (11.00) & 0.932 (5.89) & 0.922 (4.27) & & 0.980 (7.09) & 0.982 (4.84) & 0.924 (4.80) & & 0.978 (9.14) & 0.972 (5.45) & 0.926 (4.29)\\
 & 100 & 0.966 (7.47) & 0.952 (4.51) & 0.942 (3.76) & & 0.976 (3.67) & 0.980 (3.56) & 0.964 (3.46) & & 0.980 (5.40) & 0.966 (4.00) & 0.938 (3.73) \\
 & 250 & 0.970 (3.16) & 0.954 (2.82) & 0.942 (2.69) & & 0.950 (2.15) & 0.952 (2.11) & 0.952 (2.10) & & 0.954 (2.63) & 0.958 (2.51) & 0.946 (2.53) \\
\bottomrule[1.5pt]
\end{tabular} \end{sidewaystable}

\newpage

\appendix 
\section{Proofs} \label{app:proofs}
We begin by revising some standard results of kernel regression theory, which have been adapted into our context of a $2 \times 2$ contingency table. The following regularity assumptions are made:
\begin{ass} \label{ass:4diffmandf} The functions $p_{ij}(x)$, $i,j=1,2$, are bounded away from 0 and 1. Also, the marginal density of $X$, $f$, is bounded away from 0 on its compact support, $S_X$. All functions $p_{ij}(x)$ and $f$ are assumed to be four times differentiable on $S_X$. \end{ass}
\begin{ass} \label{ass:kernel} The kernel $K(\cdot)$ is a probability density function symmetric about 0 with compact support on $[-1,1]$. \end{ass}
\begin{ass} \label{ass:hbandwidth1} The common bandwidth $h = h_n$ satisfies $h \rightarrow 0$ and $nh \rightarrow \infty$ as $n \rightarrow \infty$. \end{ass}
In addition, to avoid the differing behavior kernel regression has near the boundary space of $X$ \citep{fan96}, $S_X$ is reduced to an interior support $ S^h_X = \{x \in S_X: l_X + h \le x \le u_X - h\} $ where $l_X$ and $u_X$ are the lower and upper bounds of $S_X$. Following this, we have the following adapted from \citet{wand95}:
\begin{thm} \label{thm:momentsNWest} Under Assumptions \ref{ass:multinom} and \ref{ass:4diffmandf}-\ref{ass:hbandwidth1}, it holds $\forall \, i,j = 1,2$ and $x \in S^h_X$ that
\begin{eqnarray*}
E(\widehat{p}^h_{ij}(x)) &=& p_{ij}(x) + h^2\kappa _2b_{ij}(x) + O(h^4) \\ 
Var(\widehat{p}^h_{ij}(x)) &=& \frac{\nu_0}{nhf(x)}p_{ij}(x)(1-p_{ij}(x)) + o((nh)^{-1}).
\end{eqnarray*} \end{thm}
Although already stated in the main body of the paper, we recall here the following result: if $\lim_{n\rightarrow\infty}\sqrt{nh^5} = \lambda \; \text{with} \; 0 \le \lambda < \infty$ then
\begin{equation*} 
\sqrt{nh}(\widehat{p}^h_{ij}(x) - p_{ij}(x)) \xrightarrow{d} N\left(\kappa _2\lambda b_{ij}(x), \frac{\nu_{0}}{f(x)}p_{ij}(x)(1-p_{ij}(x))\right).
\end{equation*} 
As explained in Section \ref{sec:pointwiseOR}, we treat the bias term via undersmoothing, and we thus replace Assumption \ref{ass:hbandwidth1} by
\begin{ass} \label{ass:hbandwidth2} The common bandwidth $h = h_n$ satisfies $nh^5 \rightarrow 0$ and $nh \rightarrow \infty$ as $n \rightarrow \infty$. \end{ass}
The results of Theorem \ref{thm:momentsNWest} are unchanged under this assumption, but the bias in the normality statement asymptotically vanishes and one instead obtains
$$ \sqrt{nh}(\widehat{p}^h_{ij}(x) - p_{ij}(x)) \xrightarrow{d} N\left(0, \frac{\nu_{0}}{f(x)}p_{ij}(x)(1-p_{ij}(x))\right) $$ 
and its vectorial version
\begin{equation}
\sqrt{nh}(\widehat{p}^h(x) - p(x)) \xrightarrow{d} N\left(0, \frac{\nu_{0}}{f(x)}(diag(p(x)) - p(x)p(x)^{t})\right)
\end{equation}
where $diag(p(x))$ denotes a $4 \times 4$ diagonal matrix with elements equal to the components of $p(x)$. 

\subsection{A General Expression for $Bias(\log(\widehat{OR}^h(x)))$}
We begin by evaluating $E(\log(\widehat{p}^h_{ij}(x)+\varepsilon(x)))$. To clarify, $\varepsilon(x)$ is a function of $x$ but independent of $i,j$ i.e., the same value is added to each of conditional probabilities. We also want $\varepsilon(x) \rightarrow 0$ as $n \rightarrow \infty$, since the bias of $\widehat{p}^h_{ij}(x)$ becomes negligible at large $n$ and there becomes less of a need to adjust for it. Rewriting it as follows,
\begin{equation} \label{eqn:lnphatij}
\log(\widehat{p}^h_{ij}(x) + \varepsilon(x)) = \log(p_{ij}(x)) + \log\left(1 + \frac{\widehat{p}^h_{ij}(x) + \varepsilon(x) - p_{ij}(x)}{p_{ij}(x)}\right), 
\end{equation}
then we need only consider the second term. Denoting $\widehat{\tau}_{ij}^h(x) = (\widehat{p}^h_{ij}(x) + \varepsilon(x) - p_{ij}(x))/p_{ij}(x)$, we have the following lemma regarding its moments.
\begin{lem} \label{lem:momentstau} Under Assumptions \ref{ass:multinom}, \ref{ass:4diffmandf}-\ref{ass:kernel} and \ref{ass:hbandwidth2}, it holds $\forall \, i,j = 1,2$ and $x \in S^h_X$ that if $\varepsilon(x) \rightarrow 0$ then
\begin{eqnarray*}
E(\widehat{\tau}_{ij}^h(x)) &=& \frac{1}{p_{ij}(x)}\left(\varepsilon(x) + h^2\kappa_2b_{ij}(x)\right) + o((nh)^{-1}) \\
E((\widehat{\tau}_{ij}^h(x))^2) &=& \frac{\nu_0}{nhf(x)}\frac{1-p_{ij}(x)}{p_{ij}(x)} + \frac{1}{p_{ij}(x)^2}\left(\varepsilon^2(x) + 2h^2\varepsilon(x)\kappa_2b_{ij}(x)\right) + o((nh)^{-1}) \\
E(\widehat{\tau}_{ij}^h(x)^3) & =& \varepsilon^3(x)+ 3 \varepsilon^2(x)h^2 \kappa_2 b_{ij}(x) +o((nh)^{-1}).
\end{eqnarray*} \end{lem}
\begin{prove} The first and second statements follow immediately from Theorem \ref{thm:momentsNWest}. The third moment follows from a cubic expansion $E\left(\widehat{p}^h_{ij}(x) + \varepsilon(x) - p_{ij}(x)\right)^3 = E((\widehat{p}^h_{ij}(x)-p_{ij}(x))^3) + 3\varepsilon(x) E((\widehat{p}^h_{ij}(x) - p_{ij}(x))^2) + 3\varepsilon^2(x) E(\widehat{p}^h_{ij}(x) - p_{ij}(x)) + \varepsilon^3(x)$, and utilizing the result from \citet{geenens10} that for $h = o(n^{-1/5})$, $E((\widehat{p}^h_{ij}(x) - p_{ij}(x))^4) = O((nh)^{-2})$ which implies $E(|\widehat{p}^h_{ij}(x) - p_{ij}(x)|^3) = O((nh)^{-3/2})=o((nh)^{-1}).$ \qed \end{prove}
The above result can be combined with the general formula for the log amended estimator, given by (\ref{eqn:amendedest}) in the main text, to produce the following:
\begin{lem} \label{lem:explogortilde} Under Assumptions \ref{ass:multinom}, \ref{ass:4diffmandf}-\ref{ass:kernel} and \ref{ass:hbandwidth2}, it holds for $x \in S^h_X$ that if $\varepsilon(x) \rightarrow 0$ then
\begin{align*}
E(\log(\widetilde{OR}^h(x))) &= \log(OR(x)) + h^2\kappa _2\sum\limits_{ij=11}^{22}(-1)^{i+j}\left(\frac{b_{ij}(x)}{p_{ij}(x)}\right) \\
&\quad + \, \left(\varepsilon(x)-\frac{\nu_0}{2nhf(x)}\right)\sum\limits_{ij=11}^{22}(-1)^{i+j}\left(\frac{1}{p_{ij}(x)}\right) \\
& \quad + O(\varepsilon^2(x)) + O(h^2\varepsilon(x))+ o((nh)^{-1}) + O(\varepsilon^3(x)) + O(h^2\varepsilon^2(x)). \end{align*} \end{lem}
\begin{prove} Writing $E(\log(\widetilde{OR}^h(x))) = \sum_{ij=11}^{22} (-1)^{i+j} E(\log(\hat{p}^h_{ij}(x)+\varepsilon(x)))$, then we can use (\ref{eqn:lnphatij}) to find
\begin{align} E(\log(\widetilde{OR}^h(x))) & = \sum_{ij=11}^{22} (-1)^{i+j} \log(p_{ij}(x))+\sum_{ij=11}^{22} (-1)^{i+j} E(\log(1+\widehat{\tau}_{ij}^h(x)) \notag \\
& = \log(OR(x)) +\sum_{ij=11}^{22} (-1)^{i+j} E(\log(1+\widehat{\tau}_{ij}^h(x)). \label{eqn:elog} \end{align}
Next, we apply a Taylor expansion $\log(1+\widehat{\tau}_{ij}^h(x)) = \widehat{\tau}_{ij}^h(x) - (\widehat{\tau}_{ij}^h(x))^2/2 + R(\widehat{\tau}_{ij}^h(x))$ where the remainder term can be written as
\[R(\widehat{\tau}_{ij}^h(x)) = \frac{(\widehat{\tau}_{ij}^h(x))^3}{3(1+\theta \widehat{\tau}_{ij}^h(x))^3} \]
for some $\theta \in (0,1)$. If $\widehat{\tau}_{ij}^h(x) \geq 0$, then 
\[0 \leq E(R(\widehat{\tau}_{ij}^h(x))) \leq E\left(\frac{(\widehat{\tau}_{ij}^h(x))^3}{3}\right)  \leq E((\widehat{\tau}_{ij}^h(x))^3). \]
We also know $\widehat{\tau}_{ij}^h(x) \to 0$ in probability, as $\hat{p}^h_{ij}(x)$ is a consistent estimator of $p_{ij}(x)$ and $\varepsilon(x) \to 0$. Thus, for $\widehat{\tau}_{ij}^h(x)< 0$, we can also write, provided $n$ is large enough,
\[E((\widehat{\tau}_{ij}^h(x))^3) \leq E\left(\frac{(\widehat{\tau}_{ij}^h(x))^3}{3(1+\widehat{\tau}_{ij}^h(x))^3}\right) \leq E\left(\frac{(\widehat{\tau}_{ij}^h(x))^3}{3(1+\theta \widehat{\tau}_{ij}^h(x))^3}\right)=E(R(\widehat{\tau}_{ij}^h(x))) \leq 0\]
where the first inequality holds because for $z$ negative but not too far away from 0, we have $z^3 < z^3/(3(1+z)^3)$. Hence,
\begin{equation} \label{eqn:remainder} |E(R(\widehat{\tau}_{ij}^h(x)))| \leq |E((\widehat{\tau}_{ij}^h(x))^3)| = O(\varepsilon^3(x)) + O(h^2\varepsilon^2(x))+o((nh)^{-1}) \end{equation}
as $n \to \infty$, from Lemma \ref{lem:momentstau}. Now, from the Taylor expansion, we get 
\[E(\log(1+\widehat{\tau}_{ij}^h(x))) = E(\widehat{\tau}_{ij}^h(x)) - \frac{1}{2}E((\widehat{\tau}_{ij}^h(x))^2) + E(R(\widehat{\tau}_{ij}^h(x))) \]
and using Lemma \ref{lem:momentstau} again and (\ref{eqn:remainder}) it follows
\begin{multline*} E(\log(1+\widehat{\tau}_{ij}^h(x))) = \frac{1}{p_{ij}(x)}\left(\varepsilon(x) + h^2\kappa_2b_{ij}(x)\right)  - \frac{1}{2}\frac{\nu_0}{nhf(x)}\frac{1-p_{ij}(x)}{p_{ij}(x)} \\ + O(\varepsilon^2(x)) + O(h^2\varepsilon(x))+ o((nh)^{-1}) + O(\varepsilon^3(x)) + O(h^2\varepsilon^2(x)) \end{multline*}
as $n \to \infty$. Plugging this into (\ref{eqn:elog}) yields the announced result. \qed \end{prove}

\subsection{Validity of the Multinomial-1 Bootstrap}
We begin by trying to mimic via bootstrap the asymptotic normality statement of $\widehat{p}^h_{ij}(x)$ as formulated in (\ref{eqn:phatnormalitynobias}). For the pilot bandwidth the following assumption is admitted:
\begin{ass} \label{ass:gbandwidth} The common pilot bandwidth $g = g_n$ is to be taken asymptotically larger than the optimal bandwidth $h_{opt}$, that is, $h_{opt} = o(g)$. \end{ass}
One can see that with $h_{opt} \sim n^{-1/5}$, choosing $g \sim n^{-1/9}$ as we did in the main work satisfies this. The main result of applying multinomial-1 bootstrap procedure described in Section \ref{sec:ci} is encompassed in the following theorem appropriated from \citet{rodriguez93}.
\begin{thm} \label{thm:binaryboot} Under Assumptions \ref{ass:multinom}, \ref{ass:4diffmandf}-\ref{ass:kernel}, \ref{ass:hbandwidth2}, and \ref{ass:gbandwidth}, $\forall \, z \in \mathbb{R}$, $\forall x \in S^g_X$ and $i,j = 1,2$, it holds that 
\begin{equation*}
\left| P\left(\sqrt{nh}(\widehat{p}_{ij}^{h}(x) - p_{ij}(x)) < z\right) - P^*\left(\sqrt{nh}(\widehat{p}_{ij}^{\,*h}(x) - \widehat{p}_{ij}^{g}(x)) < z\right) \right| \xrightarrow{P} 0 
\end{equation*} 
where $P^*(\cdot)$ denotes the bootstrap probability conditional on the original dataset. \end{thm} 
Note that the support for $X$ was thinned slightly from $S^h_X$ to $S^g_X = \{l_X + g, u_X - g\}$. Given $\log(\widetilde{OR}^h(x))$ is merely a continuous function of $\widehat{p}^h_{ij}(x)$, it therefore suffices to propose the following:
\begin{thm} \label{thm:bootdistamendedest} Under Assumptions \ref{ass:multinom}, \ref{ass:4diffmandf}-\ref{ass:kernel}, \ref{ass:hbandwidth2}, and \ref{ass:gbandwidth}, $\forall \, z \in \mathbb{R}$ it holds that
\begin{equation*}
\left| P\left(\sqrt{nh}\left(\log(\widetilde{OR}^h(x)) - \log(OR(x))\right) < z\right) - P^*\left(\sqrt{nh}\left(\log(\widetilde{OR}^{*h}(x)) - \log(\widetilde{OR}^g(x))\right) < z\right) \right| \xrightarrow{P} 0 \end{equation*} 
where 
\begin{align*}
\log(\widetilde{OR}^{*h}(x)) &= \log\left(\frac{(\widehat{p}^{*h}_{11}(x) + \varepsilon(x))(\widehat{p}^{*h}_{22}(x) + \varepsilon(x))}{(\widehat{p}^{*h}_{12}(x) + \varepsilon(x))(\widehat{p}^{*h}_{21}(x) + \varepsilon(x))}\right) \\
\log(\widetilde{OR}^{g}(x)) &= \log\left(\frac{(\widehat{p}^{g}_{11}(x) + \varepsilon(x))(\widehat{p}^{g}_{22}(x) + \varepsilon(x))}{(\widehat{p}^{g}_{12}(x) + \varepsilon(x))(\widehat{p}^{g}_{21}(x) + \varepsilon(x))}\right). \\
\end{align*} \end{thm}
\begin{prove} See that we can write 
\begin{align*} \log(\widetilde{OR}^h(x)) - \log(OR(x))&=\log(\widetilde{OR}^h(x)) - \log(\widehat{OR}^h(x))+\log(\widehat{OR}^h(x)) - \log(OR(x)) \\
& = \sum_{ij} (-1)^{i+j} \left\{ \log(\hat{p}_{ij}^h(x)+\varepsilon(x))-\log(\hat{p}_{ij}^h(x)) +  \log(\hat{p}_{ij}^h(x)) - \log p_{ij}(x)\right\} \\
& =\varepsilon(x) \sum_{ij}  \frac{(-1)^{i+j}}{\hat{p}_{ij}^h(x)} + \sum_{ij} (-1)^{i+j} \frac{\hat{p}_{ij}^h(x)-p_{ij}(x)}{p_{ij}(x)} \\ & \qquad \qquad +O(\varepsilon^2(x)) + O_P((\hat{p}_{ij}^h(x)-p_{ij}(x))^2)
 \end{align*}
from suitable Taylor expansions. Given $\varepsilon(x) \sim (nh)^{-1}$ and $\hat{p}_{ij}^h(x)-p_{ij}(x) =O_P( (nh)^{-1/2})$, it follows
\[\sqrt{nh}\left( \log(\widetilde{OR}^h(x)) - \log(OR(x)) \right)  =  \sqrt{nh} \sum_{ij} (-1)^{i+j} \frac{\hat{p}_{ij}^h(x)-p_{ij}(x)}{p_{ij}(x)} + O_P((nh)^{-1/2}).\]
Similarly,
\begin{align*}
 \log(\widetilde{OR}^{*h}(x)) -  \log(\widetilde{OR}^g(x)) & =  \sum_{ij} (-1)^{i+j} \log(\hat{p}_{ij}^{*h}(x)+\varepsilon(x))-\log(\hat{p}_{ij}^g(x)+\varepsilon(x)) \\
& = \sum_{ij} (-1)^{i+j} \frac{\hat{p}_{ij}^{*h}(x)-\hat{p}_{ij}^g(x)}{\hat{p}_{ij}^g(x)+\varepsilon(x)} +O_P((\hat{p}_{ij}^{*h}(x)-\hat{p}_{ij}^g(x))^2).
\end{align*}
As $\hat{p}_{ij}^g(x)+\varepsilon(x) \to p_{ij}(x)$ as $n \to \infty$ in probability, we get that the limit bootstrap distribution of $ \sqrt{nh}\left(\log(\widetilde{OR}^{*h}(x)) -  \log(\widetilde{OR}^g(x))\right)$ (i.e.\ the distribution conditional on the initial sample) is the same as the limit distribution of $\sqrt{nh}\left( \log(\widetilde{OR}^h(x)) - \log(OR(x)) \right) $, using Theorem \ref{thm:binaryboot}. \qed \end{prove}
Note the same $\varepsilon(x) = \nu_0/(2nhf(x))$ is used in the definition of $\widetilde{OR}^{g}(x)$, although a correction using the bandwidth $g$ seems more natural there. However, under Assumptions \ref{ass:hbandwidth2} and \ref{ass:gbandwidth}, $\nu_0/(2ngf(x))$ converges to 0 faster than $\nu_0/(2nhf(x))$, and the stated result is not affected by this choice. We therefore prefer using the same $\varepsilon(x)$ throughout for simplicity.

\section{Supplementary Material}
\subsection{Results of Integrated MSE for Model A using bandwidths estimated via cross-validation}
\begin{table}[H] 
\caption{\footnotesize Integrated MSE of the three nonparametric (I,II and III), the GLM, and the GAM estimators of $log(OR(x))$ for Model A at various samples sizes $n$. The best estimator is each configuration is highlighted in bold. Results for the integrated bias were very similar to original results presented in Table \ref{tab:intbias}, and therefore are not reproduced below.} \medskip \centering \label{tab:intmse2}
\begin{tabular}{lcccccc}
\toprule[1.5pt]
Model & $n$ & I & II & III & GLM & GAM \\ 
\cmidrule{3-7}
\multirow{4}{*}{A} & 50 & 1.090 & \textbf{0.494} & 0.719 & 0.893 & 1.076 \\
 & 100 & 0.517 & \textbf{0.243} & 0.366 & 0.344 & 0.646 \\
 & 250 & 0.230 & \textbf{0.113} & 0.182 & 0.127 & 0.197 \\
 & 1000 & 0.075 & \textbf{0.038} & 0.040 & 0.042 & 0.045 \\
\bottomrule[1.5pt]
\end{tabular} \end{table}

\end{document}